\documentstyle{article}
\begin{document}
\large
\begin{center}
{\LARGE Improvement of quantum key distribution protocols}\\[2.0cm]
{\large Guihua Zeng \               \  Xinmei Wang \\[0.2cm]
National Key Laboratory on ISDN of XiDian University, Xi'an, 710071, China\\[2.5cm]}
Abstract \\[0.2cm]
\end{center}

{\large The security of the previous quantum key distribution protocols,which is
guaranteed by the nature of physics law, is based on the legitimate users.
However, the impersonation of Alice or Bob by eavesdropper, in practice, will
be existed in a large probability. In this paper an improvement scheme for the
security quantum key is proposed.} 

Quantum cryptography [1] is a recently developed technique that permits two
parties, who share no secret information initially, to communicate over an
open channel and to establish between themselves a shared secret sequence of
bits. Quantum cryptography is provably secure against eavesdropping attack,
in that, as a matter of fundamental principle, the secret data can not be
compromised unknowingly to the legitimate users of the channel.
Three ingenious protocols [2-4] in quantum cryptography have been proposed. The
first, by Bennett {\it et al}, relies on the uncertainty principle of quantum
mechanics to provide key security. The security guarantee is derived from
the fact that each bit of data is encoded at random on either one of a
conjugate pair of observables of quantum-mechanical object. Because such a
pair of observables is subjected to the Heisenberg uncertainty principle [5],
measuring one of the observables necessarily randomizes the other. A further
elegant technique has been proposed by Ekert, which relies on the violation
of the Bell inequalities [6] to provide the secret security. And the third
technique, devised by Bennett, is based on the transmission of nonorthogonal
quantum states.

Raw quantum cryptography is useless in practice because limited eavesdropping
may be undetectable, yet it may leak some information, and errors are to be
expected even in the absence of eavesdropping. Also, we must protect against
an eavesdropper who would impersonate Alice for Bob and Bob for Alice. For
these reasons, quantum cryptography must be supplemented by classical tools
such as privacy amplification [7], error correction [8]. To
obtain more high security quantum privacy key, in general, four
processes has be included in the quantum key distribution: \\
a). Quantum transmission \\
b). Data sifting \\
c). Error-correction \\
d). Privacy amplification \\
Obviously, the tools from steps b) to d) are classic supplement.
For demonstration we use the quantum cryptographic protocols known as BB84
or four-state  protocols. In general, the BB84, Ekert92, and B92 protocols
possess the same process, the different is only in the method of quantum
transmission, the process is described in figure 1.

In the first step of establishing the key, Alice
sends a random sequence of signal built up from the four possibly signal
state, each appearing with equal probability. Bob possesses two measurements
apparatuses adapted to the two sets of signal states. He may distinguish
either between vertical and horizontal linear polarize photons or between
right and left circular polarize photons. For each of the signals sent to
him by Alice he chooses with equal probability an apparatuses to use. After
Bob's receiving and measurement, he sends publicly the measurement base to
Alice, and Alice compares the base between Alice and Bob's base.

After Alice and Bob obtain what is call the raw data by the quantum
transmission, the raw data must be sifted because it consists of those bits
which Bob either did not receive at all or did not correctly measure in the
basis used to transmit them. By comparison publicly the basis between Alice
and Bob, the data sifting procedure is completed.

The Third step is the data correction. A distinct feature of error correction
in quantum cryptography is that the error correction process is public, while
the transmission itself is secret. In other words, Alice and Bob must conduct
a public discussion to identify, with a high degree of confidence, all errors
in their data, while at the same time leaking as little information as possible
about the data. The basic idea is that Alice and Bob compute and exchange a
series of block check sums of their data and proceed by bi-section to locate
the error in each of the problem blocks. After their block check sums agree
several times in a row, Alice and Bob conclude that all transmission errors
have been removed. Each disclosed check sum is presumed to have been recorded by eavesdropper
Eve, and to be worth one bit of Renyi information to Eve. The number of iterations
required, and hence the amount of Renyi information leaked, depends on the desired
confidence level, the initial error rate, and the manner in which Alice and Bob
select their check sum blocks.

By the distillation art of secret key, the so called privacy amplification,
a final secure quantum key is generated and distributed. The basic principle of
privacy amplification is as follows. Let Alice and Bob shared a random
variable $W$, such as a random $n$-bit string, while an eavesdropper Eve learns a
corrected random variable $V$, providing at most $t<n$ bits of information about
$W$, i.e., $H(W|V)\leq n-t$. Eve is allowed to specify an arbitrary distribution
$P_{VW}$ (unknown to Alice and Bob) subject to the only constraint that $R(W|
V=v)\leq n-t$ with high probability (over values $v$), where $R(W|V=v)$ denotes the
second-order conditional Renyi entropy of $W$, given $V=v$. For any $s<n-t$,
Alice and Bob can distill $r=n-t-s$ bits of the secret key $K=G(W)$ while keeping
Eve's information about $K$ exponentially small in $s$ , by publicly choosing the
compression function $G$ at random from a suitable class of maps into $\{0,1\}^
{n-t-s}$.

Obviously, the above procedure is based on the legitimate users, refereed to as
Alice and Bob. However, the practice existence of impersonation of
Alice or Bob by eavesdropper, make us have to take some action to against the
eavesdropper, an efficient way is to verify the communicators' identity. 
In the follows, we improve the previous quantum key distribution scheme to
guarantee the security of quantum key for truly legitimate users .

After the privacy amplification, a compressed key is obtained, but it can not
be acted as the final key because of the impersonation. So the fifth step for
identity verification following the previous schemes, which is described in
figure 1, must be added for the security quantum key. The improving scheme is
shown in figure 2. Of course, the identity verification step can also be
inserted in the front of the privacy amplification according to the sequence:
quantum transmission $\longmapsto$ data sifting $\longmapsto$ error correcti
on $\longmapsto$ identity verification $\longmapsto$ privacy amplification, the schematic diagram is
described in figure 3. It is more practicable according to the latter
sequence, because if one of the communicators is impersonation, the
procedure may be over before the step of privacy amplification.

The key problem of the identity verification is to obtain the authentication key,
it can be established by the technique that divides the initial quantum
secret key K (it may be called Raw Key) into two parts,
i.e.,$K=K_a\oplus K_m$, where the sign $\oplus $ represents the logic plus,
the key $K_a$ is used for identity verification, while the key $K_m$ is
as a final shared secret key between Alice and Bob. The $K_a$ may be
obtained by two techniques. A single method  is to
choose the bits from K according to a proper `rule', which is adapted
publicly by users Alice and Bob,  for example, one may take the
bits in odd position in K. The guaranteed security of the
quantum key K keeps the taken bits a high degree of security, although the
`rule' is chosen publicly. Then Alice and Bob constructs
independently the authentication key $K_a$. At last Alice and Bob correct the
$K_a$ like that techniques used in the second step (Data Sifting) of quantum key distribution.
More complexity, one can adopt the privacy amplification technique again
or the hash function to obtain a shorter key as $K_a$ from the ``Raw key"
$K$. In this way, the $K_a$ is more secure and the quantum key will be not
influenced.

As shown in figure 2, after obtaining the shared dynamical-key $K_a$, Alice and Bob use it to
verify themselves identity, the technique may be like that base on the symmetric
cryptosystem. If the processes of identity verification give the `yes', the $K_m$ may act as the final key, otherwise the communication is over or re-set up. \\[2.0cm]

\begin{center}
{\bf Reference}\\[0.2cm]
\end{center}
\begin{enumerate}
\item C.H.Bennett, G.Brassard, S.Breidbart and S.Wiesner, In Advances in cryptology: Proceedings of Crypto'82, edited by D.Chaum, R.L.Rivest and A.T.Sherman (Plenum, New York, 1983).
\item C.H.Bennett,F.Bessette, G.Brassard, L.Salvail and J.Smolin, Experimental quantum cryptography, J.Cryptology 5, 3 (1992). 
\item A.K.Ekert, Quantum cyptography bases on Bell's theorem, Phys. Rev. Lett. 67, 661(1991).
\item C.H.Bennett,Quantum cryptography using any two non-orthogonal  states, Phys. Rev. Lett. 68, 3121(1992).
\item A.Bohm, quantum mechanics, Springer-Verlag New York Inc. 1979.
\item J.S.Bell, Physics (Long Island City, N.Y.) 1, 195(1965).
\item C.H.Bennett, G.Brassard, C.Crepeau and U.M.Maurer, Generalized privacy amplification, IEEE Trans. Inform. Theory, 41, 1915(1995).
\end{enumerate}
\vspace{2.0cm}

\begin{center}
{\bf Figures captions}\\[1.5cm]
\end{center}
\begin{itemize}
\item Figure 1. Schematic diagram of the quantum key distribution system.
\item Figure 2. Schematic diagram of the quantum key distribution system with identity 
verification. The identity verification step is in front of privacy amplification step a) and acts as the last step b).

\end{itemize}

\end{document}